\documentclass[showpacs,prl,twocolumn,,showpacs,paper]{revtex4}

\usepackage{epsfig}
\usepackage{amsmath}
\usepackage{graphicx}
\usepackage{dcolumn}
\usepackage{amsmath}
\usepackage{latexsym}
\usepackage{color}
\begin{document}

\title{Zero bias anomaly in a two dimensional granular insulator}
\author{N. Ossi$^1$, L. Bitton$^2$, D.B. Gutman$^1$ and A. Frydman$^1$}
\affiliation{$^{1}$ The Department of Physics, Bar Ilan University, Ramat Gan 52900,
Israel\\
$^{2}$ The Faculty of Chemistry, Weizmann Institute of Science, Rehovot 76100, Israel}

\begin{abstract}
We compare tunneling density of states (TDOS) into two ultrathin Ag films, one uniform and one granular, for different degrees of disorder.  The uniform film shows a crossover from Altshuler-Aronov (AA) zero bias anomaly to Efros Shklovskii (ES) like Coulomb gap as the disorder is increased. The granular film, on the other hand, exhibits AA behavior even deeply in the insulating regime.
We analyze the data and find  that granularity  introduces a new regime for the TDOS. While the conductivity is dominated
by hopping between clusters of grains and  is thus insulating,
the TDOS probes the properties of an individual cluster which is "metallic".
\end{abstract}

\date{\today}

\maketitle

The metal-insulator transition (MIT) in disordered electronic systems renders their transport properties one of the most exciting  topics  in  condensed matter physics.
The effects of Coulomb interaction in the vicinity of the MIT are particularly strong,
and  despite  major theoretical and experimental efforts  are not yet  fully  understood \cite{And_loc_50}.
Much of the work is focused on $2D$ metallic films. While infinite 2D systems are insulators  \cite{gof},  samples  shorter than
the localization length  $\xi$  exhibits metallic behavior. This enables to utilize thin dirty metals to study both metallic and insulating regimes.

For weakly disordered metallic films with relatively high conductivity the interplay between electron-electron interaction  and disorder  has  a pronounced  influence  on the electronic properties, as shown by Altshuler and Aronov (AA) \cite{AA2}.
Both conductivity and the tunneling density of states (TDOS)  are significantly modified (relatively to the non-interacting values)  and  acquire temperature dependent corrections. The theory predicts the depletion of the TDOS  around the Fermi level, a phenomenon that became known as zero bias anomaly (ZBA),  which was  indeed  observed experimentally in disordered metals \cite{MoBe71}.

For strongly  disordered systems,  electronic states are  localized, and transport at low temperatures is achieved via variable range hopping (VRH). For non-interacting electrons there is no gap in the TDOS at the Fermi level and the electric conductivity  is given  by Mott's law \cite{mott}.
Electron interactions open a Coulomb gap at the Fermi energy, affecting the TDOS and the electric conductivity as shown by Efros and Shklovskii (ES) \cite{ES}. The Coulomb gap in the 2D density of states is given by

\begin{equation}
\label{ES_TDOS}
\nu(\epsilon) \sim |\epsilon|\kappa^2/e^4\,,
\end{equation}
and the conductivity as a function of temperature is

\begin{equation}
\label{ES_conductivity}
\sigma(T) \simeq \exp\left(-\sqrt{\frac{T_0}{T}}\right)\,,
\end{equation}
where $\kappa$ is a dielectric constant and $T_0=e^2/\kappa \xi$.

Though the ZBA and the Coulomb gap  both result from the long range Coulomb interaction and lead to a
depletion of the TDOS at the Fermi energy,
a unified  theory of both effects is still absent \cite{Kopietz}.

\begin{figure}[tb]
{\epsfxsize=2 in \epsffile{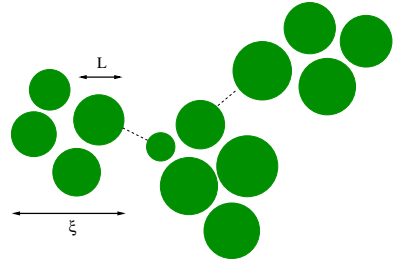}}
\vspace{0cm}
\caption{Schematic description of the system under consideration: Metallic grains of the size $L$ form clusters with typical size $\xi$. The dashed lines denote the weak links, connecting different clusters.}
\vspace{-0.5cm}
\label{grains}
\end{figure}

Most theoretical and experimental works deal with homogeneously disordered systems in which the degree of disorder controls the vicinity to the MIT.
A different system that may provide a natural and convenient arena for experimental and theoretical study of strongly interacting
systems is a granular metal \cite{Vinokur} i.e. a disordered
array of metallic islands with typical size L, embedded
in an insulating matrix (schematically shown in Fig. \ref{grains}).
In these systems the conductivity of the grains, $\sigma_0$ is high,
so that conductivity of a sample is determined by the
inter-grain coupling $g_{\rm g}$, which corresponds to the Drude
conductivity $\sigma=g_{\rm g}e^2/h$, and consequently to the diffusion
coefficient $D_{\rm g}=g_{\rm g}/\nu_0$. Clearly, the intergrain $D_g$ and $g_g$ are very different from the metallic $g_0$ and $D_0$ within the grain.

Due to the randomness in the position of the grains the coupling strengths $g_{\rm g}$  fluctuates. Deep in  the insulating regime the transport between the grains is governed by interaction-controlled VRH resulting in the ES formula,  Eq.(\ref{ES_conductivity}). Typically  $\xi > L$ so one may visualize the sample as a network of weakly connected clusters (with cluster size $\xi$).  Each cluster contains several grains, and is connected to a neighboring cluster
by  weak links with dimensionless conductivity $g_c$.
The  measured conductivity  $\sigma(\omega, q)$, is controlled
by either  $g_c \ll g_g  \ll g_0$, depending on the wave vector $q$.
In the dc limit  the conductivity of the sample is determined by $g_c$.
The TDOS, on the other hand, probes the local properties of the system, and as a function
of energy explores several regimes, as  is illustrated in Fig. 2. At energies larger than  $\epsilon_2 =e^2g_{\rm g}/2\pi C$
($C\simeq 2\pi\kappa L $ is an effective capacitance of a grain;  $t_{RC}=\hbar/\epsilon_2$ is
the corresponding  charging time of the grain) the tunneling probes the properties of the system on spatial scales smaller
than the size of a grain. Depending  on grain's  geometry,  the TDOS in this case  is the same as in the
bulk  (2D or 3D)  metal and is described  by perturbative AA theory.
For energies below
$\epsilon_1= D_{\rm c}/\xi^2$
the TDOS explores the universal low energy physics. In this limit the granular structure
of the material is irrelevant and one recovers the Coulomb gap of Eq.(\ref{ES_TDOS}) in the TDOS.

For the intermediate energy range $\epsilon_1\ll \epsilon \ll\epsilon_2$ one probes spatial distances shorter than the cluster size $\xi$, but larger than the size of a single grain. This is a new regime, that is unique to the granular material. While the dc conductivity at this energy regime is governed by tunneling between different clusters, the TDOS is determined by the properties of a single cluster. The tunneling in this regime occurs from the reference electrode into a grain,  which is
strongly coupled to other grains in the same cluster. The later can be viewed as the environment.

This kind of problem has been extensively studied in the context of the environmental ZBA in nanostructures \cite{Nazarov_book}.
It was shown that as far as TDOS is concerned,
the original problem can be mapped onto an effective electric circuit.
In this description, the microscopic details of the problem are encoded through
the  impedance of the circuit. In accordance with fluctuation-dissipation theorem this impedance
determines the spectrum of the fluctuations of the electromagnetic field.
Accounting for these fluctuation  one calculates  the  ZBA  that develops on time scales ranging between $t_{RC}$
(the time  it takes the system to screen an added electron)  and  $t_1=\hbar/\epsilon_1$
(the time at which the charged cloud reaches the boundaries of the conducting cluster).
From this consideration, it is clear that in this intermediate energy range the problem is completely equivalent to
the one solved in the context of  transport through an open
quantum dot \cite{golubev,nazarov,golubev2,Yeati,Brower}.
The tunneling conductance between the dot (analogous to the grain) and
the lead (analogous to the cluster) yields a ZBA as a function of temperature and voltage
(see Ref. \onlinecite{golubev2} for the details).
To compare the theoretical predictions  with the experimental observation we
analyze the exact results (written in terms of the digamma function) in various asymptotic limits.
Inside the narrow energy interval   $\epsilon_1 <\epsilon <  g \Delta/2\pi$
(of the order of 1K, see estimates below) the
TDOS only weakly depends on energy and may be approximated by a constant; here $\Delta$ is the energy level spacing between non-interacting single particle levels
in the dot.

For  the experimentally relevant energy range $g \Delta/2\pi \ll  \epsilon \ll \epsilon_2$ one obtains logarithmic ZBA
\begin{equation}
\label{GZ}
\nu(\epsilon)/\nu_0=1+\frac{1}{g_{\rm g}}\ln(\epsilon t_{RC})\,.  \hspace{0.5cm}
\end{equation}

The applicability  of this result for tunneling into a single open quantum dot, coupled to the environment, was  recently  confirmed experimentally in Ref. \cite{liora2}.
Note, that Eq.(\ref{GZ}) can be also viewed as the zero dimensional version of the AA theory
for short range interaction.  The presence of an external metallic tunneling electrode  makes this  model  rather realistic.

\begin{figure}[tb]
{\epsfxsize=2.7 in \epsffile{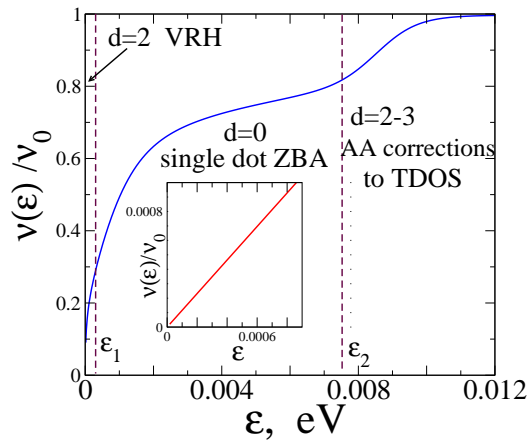}}
\vspace{0cm}
\caption{
A schematic description of the dependence of the TDOS on energy for a granular system.
In the high energy limit ($\epsilon >\epsilon_2$) one observes the TDOS  of a disordered metal,
accounted for by the  Altshuler-Aronov theory.
In the low energy limit ($\epsilon < \epsilon_1$) the system  is in the VRH regime and exhibits a  Coulomb gap given by Eq.(\ref{ES_TDOS}).
For intermediate energies ($\epsilon_1<\epsilon<\epsilon_2$)
the electric conductivity is given  by Eq. (\ref{ES_conductivity})
while the TDOS is determined by tunneling into a cluster,
Eq.(\ref{GZ}). Insert:  TDOS, given by  Eq.(\ref{ES_TDOS}), for  $\epsilon \ll \epsilon_1$.
\vspace{-0.5cm} \small}
\end{figure}

In this Letter we compare the TDOS of two thin disordered films, uniform Ag and granular Ag, as a function of $g$ and experimentally detect a regime in which Eq.(\ref{GZ}) determines the TDOS of the granular film. In this unique regime the TDOS depends logarithmically on energy despite the fact that $g$ extracted from conductivity is much smaller than unity.

The disordered 2D samples were prepared by quench condensation i.e. sequential evaporation of ultrathin layers of Ag on a cryo-cooled substrate within the measurement apparatus. This technique enables one to study the transition from insulating to metallic behavior on a \emph{single} sample without thermally cycling it or exposing it to atmosphere. Depending on the choice of substrate, one can fabricate either granular or uniform disordered films. If the samples are quench condensed on a bare SiO substrate, the resultant film is discontinuous containing separated Ag islands about 10nm in diameter \cite{kurzweil07}. Adding material results in a decrease of the average distance between the islands leading to an exponential drop of the resistance with thickness. On the other hand, when the substrate is pre-coated by a thin layer of amorphous Ge or Sb the deposited film is electrically continuous at a thickness of 1-2 monolayers of material ($\sim $5\AA ) \cite{strongin, kurzweil10} after which the Ag film obeys the usual $R_{\Box} \propto 1/d$ dependence, where $d$ is the sample thickness. These films are referred to as "uniform films". We note that one cannot entirely rule out some granularity in these films as well, however, the fact that finite conductivity is achieved for a monolayer of evaporated material followed by ohmic behavior implies that the samples are very homogeneous.

\begin{figure}[t]
    \centering
    \includegraphics[width=\columnwidth,keepaspectratio=true]{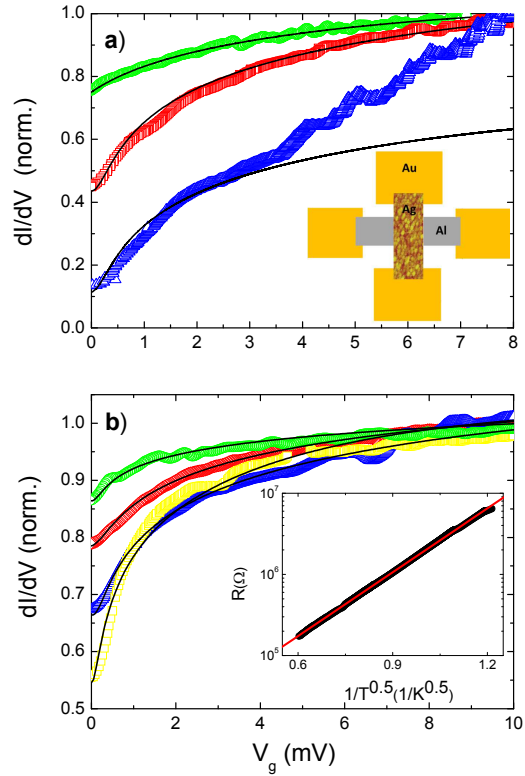}
    \vspace{-1cm}
    \caption{dI/dV versus V of uniform film having $g=13, 1, 0.35$ from top to bottom respectively (a) and a granular film having $g= 32, 8.6, 3.25, 0.2$ from top to bottom respectively (b). The solid lines are fits to Eq. \ref{GZ}. The inset in (a) shows a schematic description of the junction with a granular Ag film
as the top electrode. The inset of (b) depicts the resistance versus temperature of the most insulating stage of the granular film. The solid red line is a fit to the ES law of Eg. (\ref{ES_conductivity})  (color online).}
    \label{zba}
    \vspace{-0.5cm}
\end{figure}

For measuring TDOS we fabricated tunnel junctions using the following scheme: Prior to the cool down, four gold leads were thermally evaporated on an insulating Si/SiO2 substrate. Then a $25nm$ thick Al strip with dimensions of $1.5 mm$ by $5 mm$ was e-beam evaporated so that it connected two of the gold leads. The Al strip was then exposed to atmosphere for 2 hours thus forming an oxide layer of $Al_2O_3$ with a thickness of $3-5 nm$ that served as a tunneling barrier. The substrate was then connected to the He3 pot of a He3 fridge which was pumped to ultra high vacuum to enable film evaporation. Transport measurements were performed on the Al layer to confirm the presence of superconductivity with $T_C \sim 1.2K$. Then a $1.5 mm$ by $2 mm$ Ag strip was quench condensed either directly on the substrate for a granular film or on a pre-evaporated $2nm$ thick layer of Sb for a uniform film, thus connecting the other two gold leads as shown in the inset of Fig. \ref{zba}a.  Deposition rate and film thickness were monitored, in situ, by a quartz-crystal. The evaporation was terminated at any desired resistance. After each step of growth an R(T) measurement of the Ag film determined the degree of disorder. Then, the dI/dV versus V across the junctions were measured using standard lock-in techniques while making sure that $R_{barrier} >> R_{sample}$ so that the Ag film could be regarded as an equipotential electrode even in the insulating state. All measurements were performed at a temperature below $T_C$ of the Al. We first made sure that a BSC gap could be detected and then applied a $5 KG$ magnetic field to suppress superconductivity and measure the TDOS of the disordered film.

The differential conductance, dI/dV, representing the TDOS, as a function of voltage for both films is shown in Fig. \ref{zba} for different values of $g$. The uniform films shows a transition from a $dI/dV \propto \ln(V)$ dependence for $g=13$ (weak disorder) to $dI/dV \propto V$ for $g=0.35$ (strong disorder). This could be interpreted as a crossover from AA ZBA to an ES Coulomb-gap-like behavior. Indeed the AA theory is applicable only for $g \gg 1$, while for $g < 1$ one may expect strong localization physics to dominate. A similar crossover was reported by Butko et-al on thin films of Be \cite{butko}. It should be noted, however, that probing the Coulomb gap by a tunnel junction is problematic. On the one hand the insulating layer must be thin to allow for tunneling, but on the other hand the close electrode screens the Coulomb interaction over distances larger than the thickness of the tunneling barrier. This severely modifies the Coulomb gap which originates from long range interaction. It seems therefore that while tunneling experiments are a good tool to probe the importance of Coulomb interactions in Anderson insulators, they do not yield accurately the shape of the Coulomb gap.

The situation in the granular film is different. Fig. \ref{zba}b shows that the dI/dV curves follow a $\ln(V)$ throughout the entire crossover from weak to strong disorder. This means that even deep into the insulating regime, where $g$ extracted from conductivity is as low as 0.2 the TDOS of the granular film exhibits AA physics. The inset in Fig. \ref{zba}b shows that the conductivity in this film follows the ES law of Eq.(\ref{ES_conductivity}), so naively one would expect the TDOS to exhibit a curve more similar to the ES Coulomb gap, Eq.(\ref{ES_TDOS}). Clearly, this is not the case for the granular film. The solid lines in Fig. \ref{zba} are fits to Eq.(\ref{GZ}). It is seen that all the granular sample curves fit very well to this expression. In contrast, it was impossible to obtain a reasonable fit for the curve of the insulating uniform film ($g=0.35$).

\begin{figure}[t]
    \centering
    \includegraphics[width=\columnwidth,keepaspectratio=true]{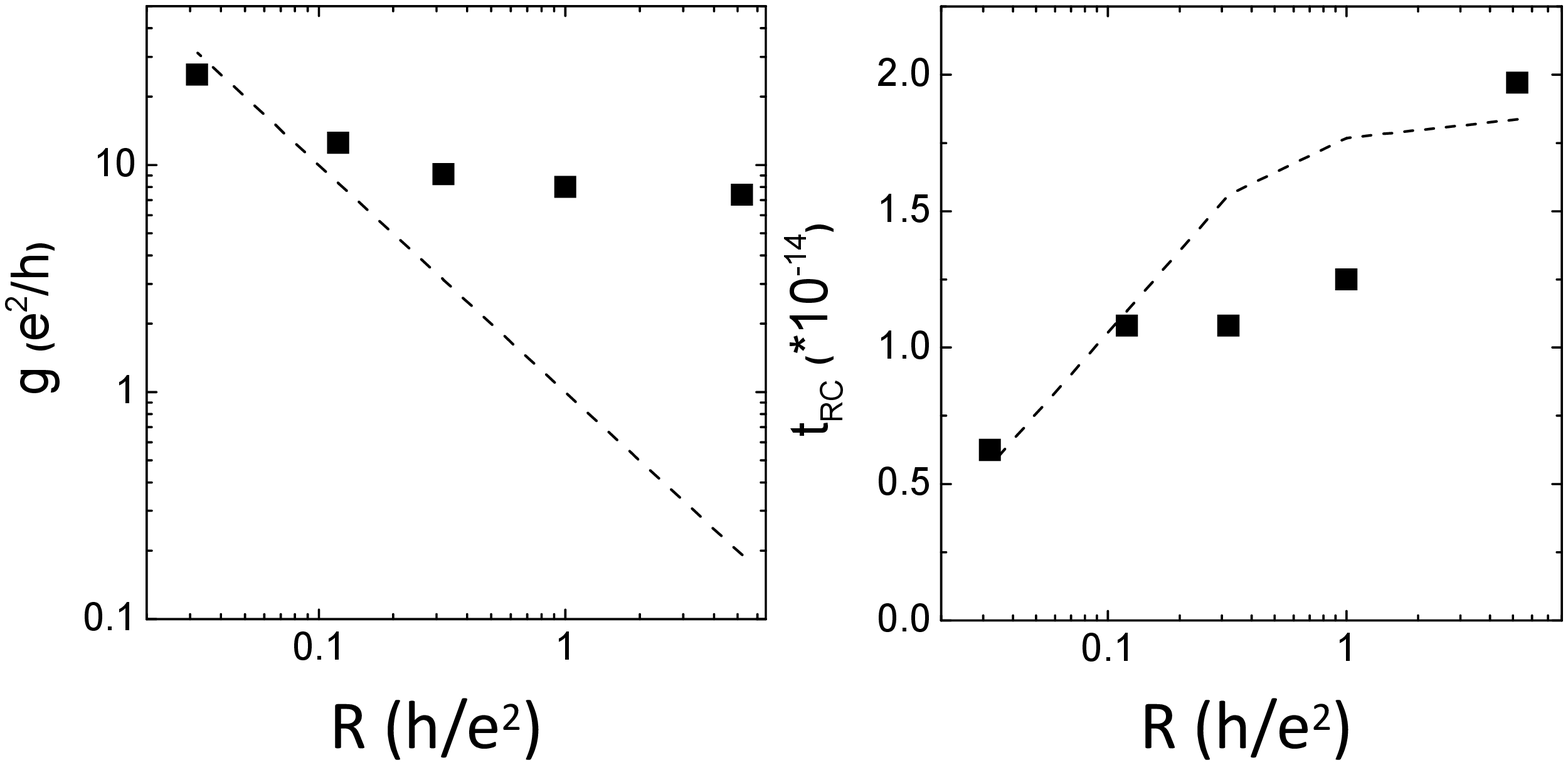}
    \vspace{-3cm}
    \caption{The dimensionless conductance, $g$, and the charging time $t_{RC}$ as a function of $R_{\Box}$ in units of $h/e^2$. Symbols are results obtained from the fits of the data in Fig. \ref{zba} (a) to Eq. \ref{GZ}. The dashed line in the left panel is the expected value from $g=1/R_\Box$. The dashed line in the right panel is the calculated value assuming a 10nm grain.}
    \label{fits}
    \vspace{-0.5cm}

\end{figure}

These fits enable one to extract  the charging time $t_{RC}$, and the conductance $g$ as a function of the disorder. Fig. \ref{fits} depicts these two fitting parameters as a function of $R_{\Box}$. The extracted $t_{RC}$ increases monotonically with $R_{\Box}$ as expected and is very close to that calculated value from Eq. (\ref{GZ}) based on the capacitance of a $10nm$ dot. This reflects the fact that, with increasing distance between grains, the electron spends more time in the cluster. On the other hand $g$ obtained from this fit is much larger than that extracted from the conductivity $g=\frac{1}{R_{\Box}}\frac{e^2}{h}$. Clearly, conductivity and TDOS probe a  different $g$. While conductivity is sensitive to the weak coupling between clusters, $g_{\rm c}$,  the TDOS apparently probes $g_{\rm g}$ which corresponds to the electrical coupling between the grains within each cluster.

Returning to the sample morphology described in Fig. \ref{grains} we note that a typical size of a grain is $L \sim 10nm$ \cite{havdala} while $\xi$ extracted from the R(T) curve of the inset of Fig. \ref{zba}b utilizing Eq.(\ref{ES_conductivity}) is found to be $\sim 100 nm$. This means that $\xi \gg L$ and the analysis presented above is appropriate. Using $g_{\rm g} \sim 10$ for metallic conductivity we get for our samples: $g\Delta/2\pi \sim 0.25 meV$ and $g e^2/2\pi\kappa C \sim 30-100 meV$.
Hence Eq.(\ref{GZ}) should hold for the gate voltage regime: $0.3mV < V < 30 mV$ . The experimental region of the gate voltage were chosen within this range, thus the TDOS is indeed expected to probe the physics of a cluster of grains and is should not be sensitive to the inter cluster weak connections.

To summarize,  we have studied  the electric conductivity and the TDOS in homogeneous and granular  films for various strengths  of disorder.  The  homogeneous  sample is  characterized by a single  energy scale, $\epsilon_1\approx D/\xi^2$,  above which the TDOS exhibits a metallic ZBA and below which there is a Coulomb gap.
This energy becomes rather large as the sample is driven deep into the insulating regime. Hence, a logarithmic dependence of the TDOS on energy will not be observed experimentally in strongly disordered homogeneous films. A granular structure  gives rise to a new energy interval, in which  ZBA is  similar to the  environmental Coulomb
 blockade  in nano-structures. This results in logarithmic dependence of TDOS on $\epsilon$ down to relatively
low energies and explains the observation of AA type ZBA in a highly disordered system.

We thank A.D. Mirlin for the discussions. This work was supported by the ISF grant numbers (399/09 and  819/10).


\begin{references}
\bibitem{gof} E. Abrahams, P.W. Anderson, D.C. Licciardello and T.V. Ramakrishnan, Phys.Rev. Lett., \textbf{42}, 673 (1979).

\bibitem{And_loc_50} for reviews on the subject see
{\it 50 Years  of Anderson Localization},
edited by Elihu Abrahams (World Scientific, Singapore, 2010)

\bibitem{AA2} B. L. Altshuler and A. G. Aronov,
in {\it Electron-Electron Interactions
in Disordered Systems}, edited by A. L. Efros and
M. Pollak (Elsevier, Amsterdam, 1985) and references therin.

\bibitem{MoBe71} for an early example see N.A. Mora, S. Bermon and J.J. Loferski,  Phys. Rev. Lett., 27, 664 (1971).
\bibitem{mott} N.F. Mott, Journal of Non- Crystalline Solids, 1, 1 (1968).

\bibitem{ES} A. L. Efros and B. I. Shklovskii, J. Phys. C \textbf{8}, L49 (1975); B. I. Shklovskii and A. L. Efros, \emph{Electronic Properties of
Doped Semiconductors} (Springer, New York, 1984).


\bibitem{Nazarov_book}
Y.V. Nazarov, Y.M. Blanter, \emph{Quantum Transport: Introduction to Nanoscience}
(Cambridge University Press, Cambridge, 2009).


\bibitem{Kopietz} for the discussion of the crossover between two regime see e.g.
P.Kopietz, Phys. Rev. Lett. {\bf 81}, 2120 (1998).


\bibitem{Vinokur}
I. S. Beloborodov, K. B. Efetov, A. V. Lopatin, and V.M. Vinokur,
Rev. Mod. Phys. {\bf 79}, 469 (2007).

\bibitem{golubev} D. S. Golubev, J. Konig, H. Schoeller, G. Schon and A. D. Zaikin,  Phys. Rev. B \textbf{56}, 15782 (1997).


\bibitem{nazarov} Y. V. Nazarov, Phys. Rev. Lett., \textbf{82}, 1245 (1999);
D. A. Bagrets and Yu. V. Nazarov Phys. Rev. Lett. {\bf 94}, 056801 (2005).

\bibitem{golubev2} D.S. Golubev and A.D. Zaikin, Phys. Rev. B {\bf 69}, 075318 (2004).


\bibitem{Yeati} A. LevyYeyati, A. Martin-Rodero, D. Esteve, and C. Urbina,
Phys. Rev. Lett. {\bf 87}, 046802 (2001).

\bibitem{Brower} P. W. Brouwer, A. Lamacraft, K. Flensberg, Phys. Rev. B {\bf 72}, 075316 (2005).


\bibitem{liora2} L. Bitton, D.B. Gutman, R. Berkovits, and A. Frydman,
Phys. Rev Lett. {\bf 106}, 016803 (2011).

\bibitem{kurzweil07} N. Kurzweil and A. Frydman, Phys. Rev B., rapid communications, 75 020202(R) (2007).
\bibitem{strongin} M. Strongin, R. Thompson, O. Kammerer and J. Crow, Phys.
Rev. {\bf B1}, 1078 (1970).
\bibitem{kurzweil10} N. Kurzweil, E. Kogan and A. Frydman, Phys. Rev. B, 82, 235104 (2010).

\bibitem{butko} V.Y. Butko, J.F. DiTusa,  and P.W. Adams, Phys. Rev. Lett. 84, 1543 (2000).

\bibitem{havdala} T. Havdala, A. Eisenbach and A. Frydman, unpublished.


\end{references}
\end{document}